# Machine Learning Relationships between Nanoporous Structures and Electrochemical Performance in MOF Supercapacitors

Zhenxiang Wang, Taizheng Wu, Liang Zeng, Jiaxing Peng, Ding Yu, Ming Gao, and Guang Feng*

Zhenxiang Wang, Taizheng Wu, Liang Zeng, Jiaxing Peng, Ding Yu, Ming Gao, and Guang Feng

State Key Laboratory of Coal Combustion, School of Energy and Power Engineering, Huazhong University of Science and Technology, Wuhan 430074, China.

Guang Feng

Institute of Interdisciplinary Research for Mathematics and Applied Science, Huazhong University of Science and Technology, Wuhan 430074, China

Corresponding author's email: gfeng@hust.edu.cn

**ABSTRACT:** The development of supercapacitors is impeded by the unclear relationships between nanoporous electrode structures and electrochemical performance, primarily due to challenges in decoupling the complex interdependencies of various structural descriptors. While machine learning (ML) techniques offer a promising solution, their application is hindered by the lack of large, unified databases. Herein, we use constant-potential molecular simulation to construct a unified supercapacitor database with hundreds of metal-organic framework (MOF) electrodes. Leveraging this database, well-trained decision-tree-based ML models achieve fast, accurate, and interpretable predictions of capacitance and charging rate, experimentally validated by a representative case. SHAP analyses reveal that specific surface area (SSA) governs gravimetric capacitance while pore size effects are minimal, attributed to the strong dependence of electrode-ion coordination on SSA rather than pore size. SSA and porosity, respectively, dominate volumetric capacitance in 1D-pore and 3D-pore MOFs, pinnacling the indispensable effects of pore dimensionality. Meanwhile, porosity is found to be the most decisive factor in the charging rate for both 1D-pore and 3D-pore MOFs. Especially for 3D-pore MOFs, an exponential increase with porosity is observed in both ionic conductance and in-pore ion diffusion coefficient, ascribed to loosened ion packing. These findings provide profound insights for the design of high-performance supercapacitor electrodes.

# 1. Introduction

Electrical double-layer capacitors (EDLCs), known as supercapacitors, are a promising class of energy storage devices that offer high power density and long cycling life but moderate energy density[1, 2]. Enhancing energy density requires the development of nanoporous electrode materials[3, 4], as substantial efforts have been devoted to tuning pore structures to improve their capacitive performance[5-7]. Extensive experimental and computational studies have demonstrated a strong dependence of capacitance on electrode pore size and specific surface area (SSA)[8-11]. However, constant-potential molecular dynamics (MD) simulations unveiled no definitive correlation between capacitance and structural descriptors such as pore size, porosity, or density, based on 19 ordered porous electrodes of zeolite-templated carbons[12]. Very recent experiments of nuclear magnetic resonance spectroscopy on 20 amorphous nanoporous carbons suggest that the pore size and SSA have weak impacts on capacitance[13]. These different findings underscore the necessity for better understanding relationships between electrode structure and electrochemical performance, which remains a challenge primarily due to difficulties in accurately characterizing electrode topology[14, 15] and disentangling the combined contributions of various structural descriptors.

Machine learning (ML) has proven highly effective in investigating intricate structure-performance relationships, requiring a large database that pairs well-defined material structures with corresponding electrochemical performance metrics[16, 17]. Collecting hundreds of porous carbon structures with capacitive performance data from literature reports, data-driven studies have utilized ML models to predict capacitance[18-22]. All of these models emphasize the importance of SSA in determining capacitance, while some of them consider pore size as an important indicator[18, 20] and others can predict capacitance regardless of pore size[21, 22]. This difference may come from the non-uniformity in structural descriptors from un-unified experiments, because pore structures of amorphous carbons are inherently complicated and hard to charterize[23, 24]. In particular, the structural disorder of porous carbons, which significantly affects gravimetric capacitance[13], is challenging to quantify and absent from these databases. In contrast, metal-organic frameworks (MOFs), with well-defined and ordered structures that facilitate nanopore characterization[25], provide an ideal platform for mechanism exploration[26]. High-throughput density functional theory (DFT) calculations on MOFs have successfully mapped crystal structures to bulk properties of band gap[27] and conductivity[28], while a systematic study on electrochemical interfaces is still unexplored.

In this work, we aim to dissect the relationships between porous electrode topology and electrochemical performance by building a large and unified database of MOF supercapacitors. Constant-potential MD simulations are employed to calculate electrochemical performance and probe the nanoscale interface, as this approach can effectively capture both equilibrium states and dynamic processes in supercapacitors based on various nanoporous materials[29-31]. Taking 20,375 DFT-optimized MOF structures[27, 32], we developed a high-throughput simulation workflow, obtaining a unified database composed of 510 EDLCs with MOF electrodes and ionic liquid electrolyte. By training gradient-boosted decision tree models on this MD database, we predict the capacitance and charging rate of 3740 additional porous MOF EDLCs, with experimental validation for a representative case. Through interpretable techniques, we assessed the individual contribution of diverse pore structure descriptors to capacitance and charging dynamics. Results illuminate that SSA determines gravimetric capacitance; SSA and porosity, respectively, govern volumetric capacitance of EDLCs based on MOFs with 1D and 3D pores; porosity is the primary determinant of charging dynamics in both 1D-pore and 3D-pore MOFs. Furthermore, the charge storage mechanism is elucidated by examining ion-electrode coordination, and an exponential increase with porosity is observed in both ionic conductance and in-pore ion diffusivity in 3D-pore MOFs.

## 2. Results and Discussion

### 2.1 Supercapacitor database construction

We chose 20,375 DFT-optimized MOF crystal structures, adopting the QMOF database[27, 32], as candidate electrode materials. Pore structure analysis, based on Voronoi decomposition[33], is performed for all crystals. Six structural descriptors are used to characterize the nanostructures of MOFs (**Figure 1a** and **Figure S1**), including pore dimensionality, pore limiting diameter (PLD), largest cavity diameter (LCD), SSA, porosity ($\epsilon_{void}$), and density ($\rho$). A subset of 4250 MOFs with SSA exceeding a typical value for porous MOFs (400 m$^2$ g$^{-1}$)[34] is selected as supercapacitor electrodes. Constant-potential MD simulations are performed to acquire their electrochemical performance. Specifically, for each porous MOF, a symmetric EDLC, consisting of two identical

MOF electrodes immersed in 1-ethyl-3-methylimidazolium tetrafluoroborate ([EMIM][BF$_4$]) electrolyte, is set up under voltages of 0-4 V (details see Methods and Section 1 of SI including **Figure S2** and **Table S1**). Gravimetric capacitance ($C_g$) and volumetric capacitance ($C_v$) are calculated based on the charge storage under the equilibrium state. Since all the EDLCs are charged under constant voltage, the charging rate, calculated by the average current density per unit mass ($I$) during charging is calculated to assess power performance. As a result, 510 MOF EDLCs are stored in the EDLC database, with each entry containing six structure descriptors for the MOF electrode and three electrochemical performance metrics (**Figure 1a**). It is worth noting that although not all of the MOFs are intrinsically conductive, their conductivity could be potentially enhanced by doping or functional group modification, with minimal changes to the structure[35, 36].

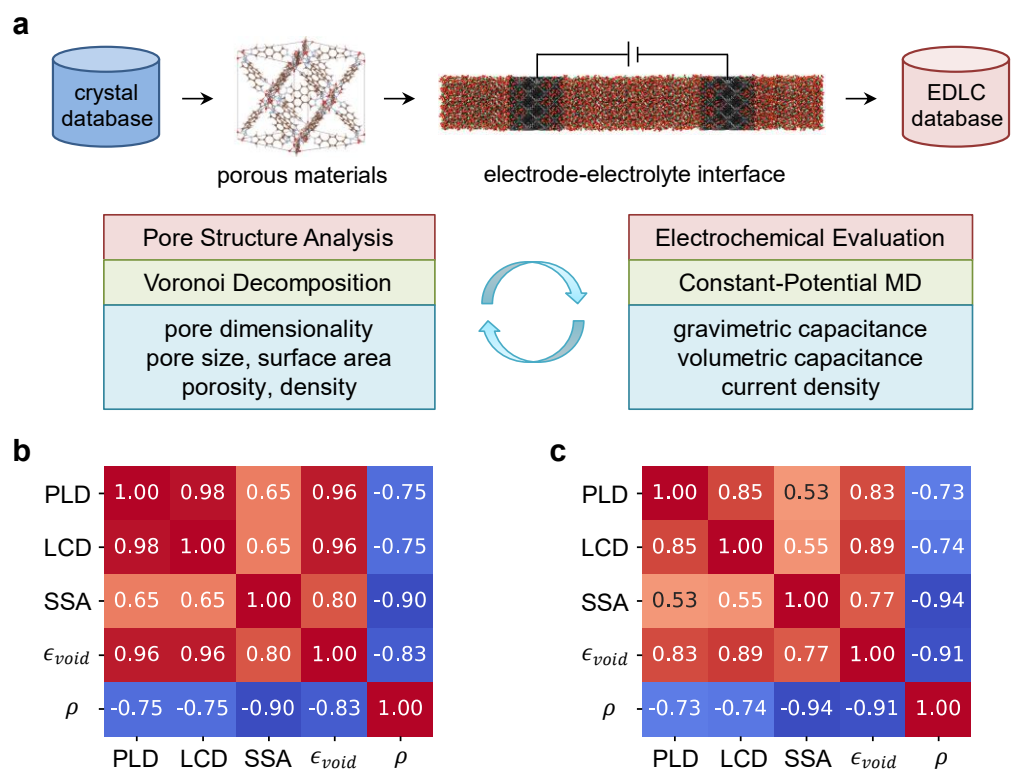

**Figure 1. Schematic of the workflow and Spearman correlation analysis heat maps. a,** Pore structure analysis is performed on crystals from the QMOF database, with porous materials selected for electrochemical evaluation based on constant-potential MD. Only ion-accessible porous MOFs are included in the EDLC database. **b-c,** Spearman rank correlation coefficients between each structural descriptor of 1D-pore (**b**) and 3D-pore (**c**) MOFs.

Regarding pore dimensionality as a classification indicator, only 12 MOFs with 2D pores are recorded in the EDLC database, and then we mainly compare 1D-pore (322 entries) and 3D-pore (176 entries) MOFs in this work. 3D-pore MOFs demonstrate larger LCD (up to 3.1 nm), higher

SSA (up to 7370 m$^2$ g$^{-1}$), greater porosity (up to 0.87), and lower density (down to 0.087 g cm$^{-3}$), compared to 1D-pore MOFs with the same PLD (**Figure S3**). The Spearman correlation analysis is employed for all MOFs in the EDLC database, showing strong correlations between six structural descriptors, with the majority of absolute correlation coefficients exceeding 0.8 (**Figure 1b,c**).

## 2.2 Machine learning predictions and interpretation

### 2.2.1. Predictions of capacitance and charging dynamics

The strong correlations between structural descriptors make traditional data analysis methods unavailable to isolate the effects of individual descriptors on electrochemical performance[37, 38]. Therefore, an ML model, using XGBoost[39], is adopted to map the structural descriptor vectors to EDLC performance metrics. Shapley additive explanations (SHAP), which quantify the marginal contribution of each variable to ML predictions[40, 41], are utilized to interpret non-linear structure-performance relationships and evaluate contributions of each structural descriptor to electrochemical performance (details see Methods and Section 2 of SI including **Table S2**). We took 80% of structures to train ML models for predicting EDLC performance and reserved the remaining 20% for model evaluation. For gravimetric capacitance and current densities (**Figure 2a,e**), nearly all data points lie close to the perfect prediction line, with R$^2$ values around 0.9 on the test set. For volumetric capacitance (**Figure 2c**), the prediction is reliable for dependence analysis, with R$^2$ values of 0.95 and 0.80 on the training and test sets, respectively.

To further validate our ML model in predicting capacitance, we chose a representative MOF, Ni$_3$(2,3,6,7,10,11-hexaiminotriphenylene)$_2$ (Ni$_3$(HITP)$_2$)[26, 31, 42], which is not included in either QMOF or EDLC database. We synthesized high-crystallinity Ni$_3$(HITP)$_2$ with an SSA of 1051 m$^2$ g$^{-1}$, very close to the theoretical value of 1091 m$^2$ g$^{-1}$ (details see Methods and Section 3 of SI with **Figure S4-S7**). The ML-predicted capacitance is 102 F g$^{-1}$, in good agreement with the experimental measurement of 116 F g$^{-1}$, confirming the reliability of this ML model in predicting electrochemical performance (details see Section 4 of SI with **Table S3**). Electrochemical performance has been predicted for the other 3740 porous MOFs without MD calculations. 50 promising high-performance MOF electrodes are found with gravimetric capacitance over 215 F g$^{-1}$ or volumetric capacitance exceeding 130 F cm$^{-3}$ (available in Supporting Table). The highest predicted gravimetric capacitance is 227 F g$^{-1}$, corresponding to an energy density of 126

Wh kg$^{-1}$, very competitive among electrode materials for EDLCs with the highest energy densities reported in the last ten years (**Table S4**). Importantly, such an MOF could attain quite rapid charging, about 6 times faster than Ni$_3$(HITP)$_2$, as Ni$_3$(HITP)$_2$ renders comparable power density to porous carbon electrodes[31].

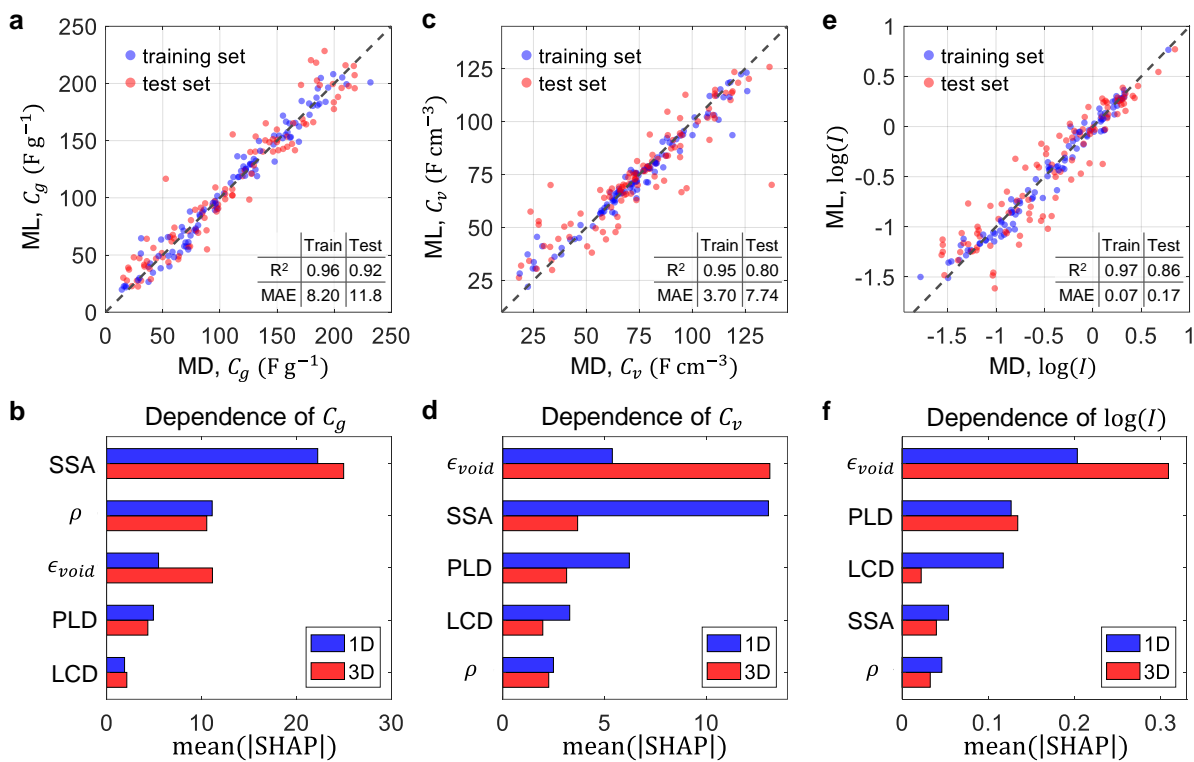

**Figure 2. Prediction of EDLC performance based on structural descriptors and importance of each descriptor.** Predictions and dependences of gravimetric capacitance (**a-b**), volumetric capacitance (**c-d**), and logarithmic current density (**e-f**). The dashed grey line represents perfect prediction.

*2.2.2. Importance of structural descriptors*

The importance of structural descriptors in determining electrochemical performance is ranked based on the mean of absolute SHAP values. For gravimetric capacitance (**Figure 2b**), SSA emerges as the most decisive factor of EDLCs; whereas the marginal contribution of pore size is minimal. This finding is different from prior studies that either emphasize the importance of both SSA and pore size[8-10] or conclude that both have a weak impact on gravimetric capacitance[12, 13]. For volumetric capacitance (**Figure 2d**), SSA is the most critical descriptor for 1D-pore MOFs, while in 3D-pore MOFs, porosity is the dominant factor. This contrast suggests that pore dimensionality is of great significance in determining volumetric capacitance. Notably, SHAP analysis exhibits that pore size effects are minimal, though previous researches report pore size as an important factor in volumetric capacitance[43]. Regarding charging dynamics, previous studies detected a primary role of pore size[9, 31, 44-46]; however, **Figure 2f** illustrates that porosity is the principal determinant of current density in both 1D-pore and 3D-pore MOFs, and pore size contribution is minor, even subtle for 3D-pore MOFs.

**2.3 Molecular insights into structure-performance relationships**

*2.3.1. Gravimetric capacitance*

The SHAP summary plots are obtained to scrutinize effects of structural descriptors on the gravimetric capacitance. Both PLD and LCD exhibit a narrow distribution near a SHAP value of zero (**Figure 3a,b**), confirming the weak correlation between pore size and gravimetric capacitance (**Figure 2b,3c** and **Figure S8a** in Section 5 of SI). In contrast, the SSA shows a wide distribution of SHAP values, highlighting its significant impact and demonstrating a strong positive correlation with gravimetric capacitance (**Figure 3a,b**). The maximum gravimetric capacitance could be achieved of 231 F g$^{-1}$ for 1D-pore MOFs with an SSA of 2693 m$^2$ g$^{-1}$ and 256 F g$^{-1}$ for 3D-pore MOFs with an SSA of 6189 m$^2$ g$^{-1}$ (**Figure 3d**). It is worth noting that the marginal contribution of porosity is negative in 3D-pore MOFs according to SHAP analysis (**Figure 3b**), although a seemingly positive correlation between porosity and gravimetric capacitance can be found if other structural descriptors are not controlled (**Figure S8b**). This underlines the necessity of utilizing interpretable ML models to disentangle the coupled effects of structural descriptors on electrochemical performance.

To investigate the molecular origin of SSA-determined gravimetric capacitance, the average coordination number of counter-ions (cations in the positive electrode and anions in the negative electrode) surrounding each electrode atom ($CN_{anion}^{pos}$ and $CN_{cation}^{neg}$) is analyzed to reflect counter-ion adsorption ability (**Figure S9**). For both positively and negatively polarized MOF electrodes, the number of coordinated counter-ions increases with SSA (**Figure 3e** and **Figure S10a**), as more electrode atoms are exposed to the electrolyte. The coordination number of anions is found to be higher than that of cations due to the smaller size of BF$_4^-$ than EMIM$^+$ (**Figure 3e** vs. **Figure S10a**). Remarkably, a strong positive linear relationship is found between the coordination number of counter-ions and the average charge stored per electrode atom ($Q_{atom}$), proving that coordinated counter-ions play a critical role in charge storage (**Figure 3f** and **Figure S10b**). Consequently, $Q_{atom}$ increases with SSA (**Figure S11a**). The charge stored in the electrode per unit mass is analyzed by the product of $Q_{atom}$ and the number of electrode atoms per unit mass ($\rho_{atom}^{mass}$), where $\rho_{atom}^{mass}$ depends solely on the chemical composition of electrode materials. For the MOF materials studied, $\rho_{atom}^{mass}$ is found to be insensitive to SSA (**Figure S11b**). Therefore, $Q_{atom}$ increases with SSA while $\rho_{atom}^{mass}$ remains relatively constant, leading to enhanced gravimetric capacitance. In contrast, the coordination number of counter-ions is weakly linked to pore size (**Figure S10c,d**), explaining the minor influence of pore size on gravimetric capacitance.

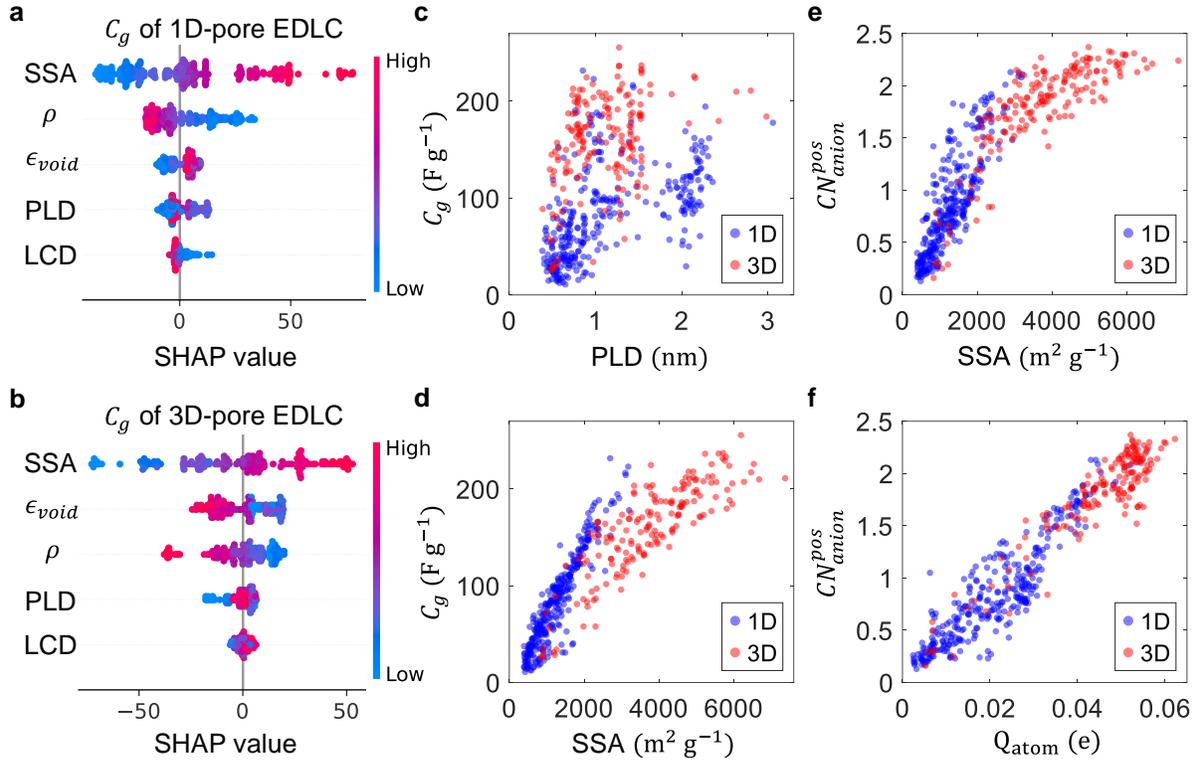

**Figure 3. Dependence of gravimetric capacitance. a-b,** SHAP analysis summary plots for gravimetric capacitance of 1D-pore (**a**) and 3D-pore (**b**) EDLCs. **c-d,** Correlations between gravimetric capacitance and PLD (**c**) and SSA (**d**). **e-f,** Correlations between average coordination number of counter-ions around electrode atoms and SSA (**e**) and the magnitude of electrode charge per atom (**f**).

*2.3.2. Volumetric capacitance*

The SHAP summary plots of volumetric capacitance (**Figure 4a,b**) consolidate **Figure 2d** that the most important factor is SSA for 1D-pore MOFs and porosity for 3D-pore MOFs. For 1D-pore MOFs, a positive correlation exists between volumetric capacitance and SSA (**Figure 4c**). Volumetric capacitance exhibits a wide distribution (16 - 138 F cm$^{-3}$) for subnanometer pores with small porosity, and the distribution becomes narrower (56 - 80 F cm$^{-3}$) as pore size and porosity increase (**Figure 4d** and **Figure S12**). This indicates that SSA is a better indicator than porosity or pore size. For 3D-pore MOFs, neither SSA nor pore size shows significant effects on volumetric capacitance (**Figure 4b** and **Figure S7**). As porosity gets enlarged, volumetric capacitance initially rises and then declines. This volcano-shaped relation, with the highest volumetric capacitance of 138 F cm$^{-3}$, is consistent with previous experimental observations of EDLCs based on ultra-thick graphene electrodes[6].

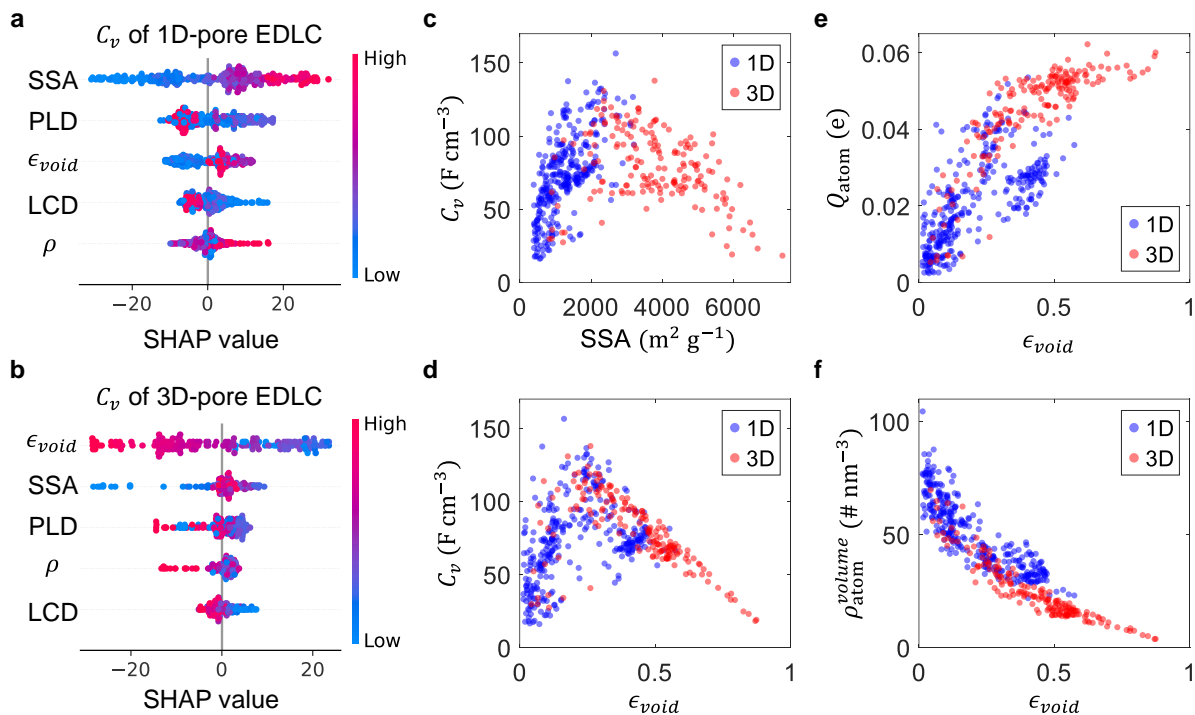

**Figure 4. Dependence of volumetric capacitance. a-b,** SHAP analysis summary plots of volumetric capacitance for 1D-pore (**a**) and 3D-pore (**b**) EDLCs. **c-d,** Correlations between volumetric capacitance and SSA (**c**) and porosity (**d**). **e-f,** Correlations between porosity and magnitude of electrode charge per atom (**e**) and volumetric number density of electrode atom (**f**).

To unveil the molecular mechanism underlying the dependence of volumetric capacitance on SSA and porosity, the charge stored in the electrode per unit volume is analyzed by the product of $Q_{atom}$ and the number of electrode atoms per unit volume ($\rho_{atom}^{volume}$). For 1D-pore MOFs, $Q_{atom}$ shows a wide distribution (from 0.002 to 0.04 e) at low porosity, consistent with the wide distribution of SSA at a given porosity (**Figure S3e**). As porosity increases, the lower limit of $Q_{atom}$ increases to ~0.02 e with the growing lower limit of SSA, and the upper limit varies less (**Figure 4e**). Meanwhile, $\rho_{atom}^{volume}$ decreases monotonically with increasing porosity (**Figure 4f**). Therefore, as porosity rises, the lower limit of volumetric capacitance gets enlarged owing to the increased lower limit of $Q_{atom}$, while the upper limit decreases due to the decline in $\rho_{atom}^{volume}$, accounting for the narrower distribution. Since counter-ion coordination is moderately affected by porosity and strongly dependent on SSA, SSA dominates $C_v$ of 1D-pore MOFs. For 3D-pore MOFs, $Q_{atom}$ surges with porosity as a result of more coordinated counter-ions, and then grows slightly when porosity exceeds 0.25 (**Figure 4e**). $\rho_{atom}^{volume}$ of 3D-pore MOFs also exhibits a negative correlation with porosity (**Figure 4f**). As a result, with increasing porosity, the volumetric capacitance initially rises, thanks to the fast growth of coordinated counter-ions and $Q_{atom}$, and it subsequently decreases because the declining $\rho_{atom}^{volume}$ becomes the dominant factor in volumetric capacitance. This transition highlights a careful choice of porosity when optimizing volumetric capacitance for 3D-pore electrodes.

*2.3.3. Charging dynamics*

We then concentrate on the current density that reflects the power performance of EDLCs. SHAP summary plots confirm that porosity determines the current density of both 1D-pore and 3D-pore MOFs (**Figure 5a,b**). Although a positive correlation is found between current density and PLD (**Figure 5c**), for 1D subnanometer pores, the current density exhibits a wide distribution at a given PLD, indicating that pore size is a weak indicator for such pores. Porosity proves to be a more reliable descriptor than pore size for both 1D-pore and 3D-pore MOFs, because of narrower distributions of current density versus porosity (**Figure 5d**). Especially for 3D-pore MOFs, a highly linear relationship is found between $\log(I)$ and porosity, indicating that current density grows exponentially with porosity. Therefore, the ignorance of porosity effects when tuning pore size could account for the different findings in previous studies reporting either a monotonical[9, 45, 46] or oscillatory[44, 47] increase in charging dynamics with larger pore size.

To examine how high porosity enhances charging dynamics, we analyzed in-pore ion diffusivity. For 1D-pore MOFs at 0 V, the diffusion coefficient loosely grows with porosity (Figure 5e), consistent with the accelerated charging dynamics (**Figure 5d**). Interestingly, anomalously fast diffusion is found in some subnanometer pores (**Figure 5e** and **Figure S13a**); however, the fast ion diffusion in these MOFs does not result in high current density (**Figure 5d** and **Figure S13b**). This suggests a breakdown of the Nernst-Einstein relation describing a linear growth of ionic conductance with diffusion (**Figure 5f**)[48], which may originate from the additional steric hindrance[30, 44] during the charging process. For 3D-pore MOFs, the diffusion coefficient is found to increase exponentially with porosity (**Figure 5e**), accounting for the observed exponential current-porosity relationship (**Figure 5d**). The enhanced diffusion coefficient ensures faster charging, as a positive relation is found between diffusion and current density (**Figure 5f**). We further calculated in-pore ion density ($\rho_{ion}$, the ratio of ion mass to pore volume) to investigate the dependence of diffusion coefficient on porosity. Results show a strong negative correlation between ion diffusion and ion density (**Figure S14**). As porosity gets enlarged, the in-pore ion density decreases due to the weakened nanoconfinement, leading to faster ion diffusion and then charging dynamics. The different diffusion-current relationships for pores with different dimensionalities demonstrates better ion transport during charging 3D-pore MOFs.

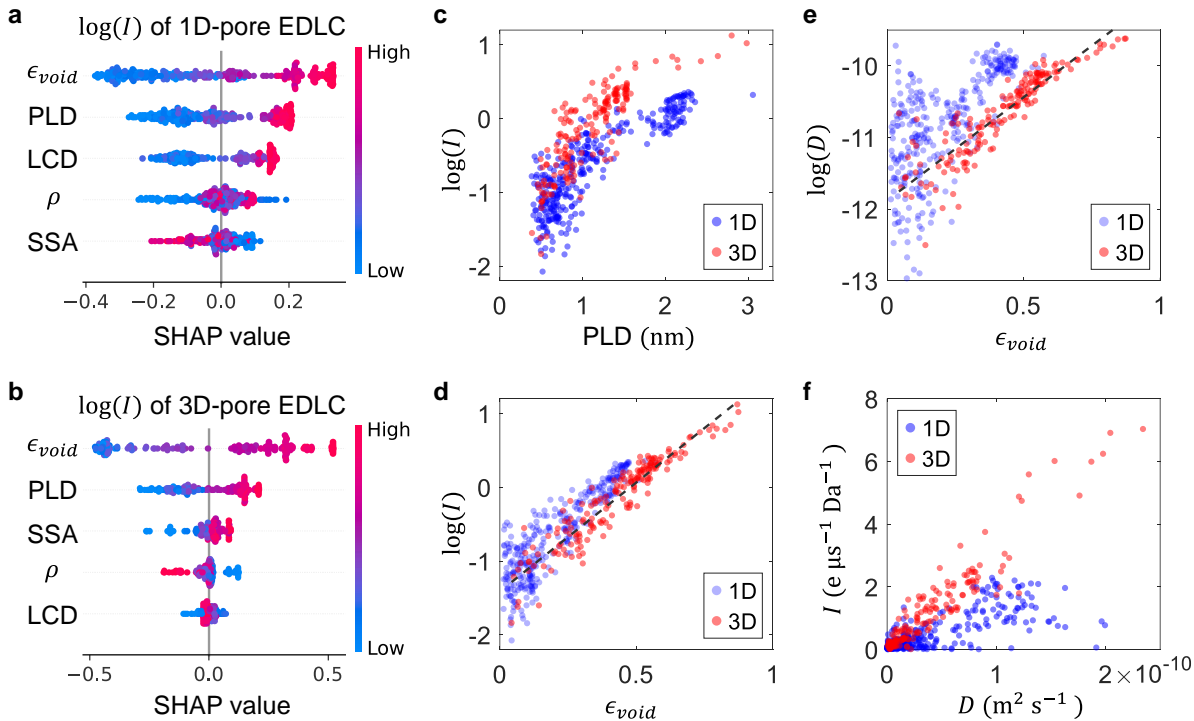

**Figure 5. Dependence of charging dynamics. a-b,** SHAP analysis summary plots for logarithmic current density of 1D-pore (**a**) and 3D-pore (**b**) EDLCs. **c-d,** Correlations between current density and PLD (**c**) and porosity (**d**). **e,** Correlation between in-pore ion diffusion coefficient and porosity. **f,** Correlation between current density and in-pore ion diffusion coefficient.

## 3. Conclusion

In summary, a large and unified EDLC database is constructed using high-throughput constant-potential MD simulations on hundreds of MOF electrodes. Through training decision-tree-based models on this database, we achieved rapid, accurate, and interpretable predictions of both capacitance and charging rate, validated experimentally by a representative MOF. Utilizing SHAP analyses, we disentangled the combined contributions of six structural descriptors to capacitance and charging dynamics. Results demonstrate that SSA is the primary determinant of gravimetric capacitance while the effects of pore size are subtle, different from previous studies that regard both SSA and pore size as either important indicators[8-10] or weak parameters[12, 13]. It can be explained by the finding that the counter-ion coordination to electrode strongly depends on SSA rather than the pore size. For volumetric capacitance, SSA is identified as the most influential factor for 1D-pore MOFs, whereas porosity dominates for 3D-pore MOFs, suggesting that pore dimensionality plays a crucial role in determining volumetric capacitance. Effects of pore size are found to be minor in MOFs, unlike previous studies attaching importance to pore size in determining the volumetric capacitance of porous carbons[43]. Regarding charging dynamics, porosity determines the current density of both 1D-pore and 3D-pore MOFs, while previous studies highlight a strong dependence of charging dynamics on pore size[9, 31, 44-46]. Notably, for 3D-pore MOFs, both current density and in-pore ion diffusion coefficient are found to increase exponentially with porosity, attributed to loosened in-pore ion packing. This relationship requires further experimental validation.

Although this work focuses on MOF electrodes in ionic liquid electrolyte, the insights may inform the structure-performance relationships for other nanoporous materials, such as covalent organic frameworks[49], zeolite-templated materials[50], and their derivatives[51]. Through techniques like transfer learning[52], the knowledge derived from this database of MOF supercapacitors could be extended to other electrode-electrolyte combinations with fewer samples in need. Coupling simulation accuracy with ML efficiency, this work establishes a framework for understanding effects of electrode structure on electrochemical performance, paving the way for the design and optimization of high-performance energy storage materials.

## 4. Methods

### 4.1 MD simulation

Each MD simulation system contains two identical symmetric electrodes and an appropriate number of electrolyte molecules to ensure that the electrolyte density in the central region of the reservoir remains in a bulk-like state at a voltage of 0 V (deviation within 1%). The electrode separation is 20 nm. After the application of a jump-wise voltage, the total charge on the positive electrode was averaged every 5 ns. Equilibrium was considered achieved when variations in charge quantity between consecutive records were within 0.5%. At the equilibrium state under polarization, the electrolyte density in the central region of the reservoir was rechecked, and data were excluded if the bulk density deviated by more than 5% from the bulk-like state. Additionally, electrodes were classified as ion-inaccessible if no ions entered the electrode pores before and after charging. Such electrodes were excluded from the EDLC database.

The Lennard-Jones parameters for the MOF atoms were taken from the generic universal force field[53]. An all-atom virtual site model[54], which could provide proper thermodynamic and dynamic properties including density, diffusion coefficient, and electrical conductivity, was adopted for [EMIM][BF$_4$] (**Table S1**). Periodic boundary conditions were applied in all directions. All the MD simulations are performed in the NVT ensemble using a customized MD software GROMACS[55]. The applied electrical potential between the two electrodes was maintained by the constant-potential method, as it allows the fluctuations of charges on electrode atoms subject to an equipotential constraint during the simulation[56, 57]. To guarantee accuracy, the electrode charges are updated on the fly at every simulation step (2 fs). The electrolyte temperature was maintained at 400 K using the V-rescale thermostat[58]. The in-pore ion diffusion coefficient along the z-direction (**Figure S2**) under 0 V was calculated through mean square displacement. The electrostatic interactions were computed using the particle mesh Ewald method[59]. A cutoff length of 1.2 nm was used in the direct summation of the non-electrostatic interactions and electrostatic interactions in real space.

**4.2 Machine learning**

A gradient-boosted decision tree ML model, implemented in XGBoost[39], was used to map the structural descriptor vectors to EDLC performance metrics. Specifically, six structure descriptors, including channel dimensionality, PLD, LCD, SSA, porosity, and density, were used to predict three EDLC performance metrics: gravimetric capacitance ($C_g$), volumetric capacitance ($C_v$), and logarithmic current density, $\log(I)$. Due to the wide range of current densities, spanning three orders of magnitude, their logarithmic values are used as the fitting target to minimize prediction bias. The dataset was split, such that 80% of materials were used for training and the rest 20% were reserved for testing. Hyperparameters of the gradient-boosted decision tree model (listed in **Table S2**) were optimized using tenfold cross-validation to enhance generalization and minimize overfitting.

### 4.3 Experiments of $Ni_3(HITP)_2$ synthesis

Starting materials were purchased from commercial sources and used without further purification unless otherwise noted. Especially, $Ni(OAc)_2·4H_2O$ and NaOAc were purchased from Aladdin; 2,3,6,7,10,11-hexaaminotriphenylene hexahydrochloride (HATP·6HCl) was purchased from Yanshen Technology Co., Ltd.; N,N-dimethylformamide (DMF), N,N-dimethylacetamide (DMA) and methanol were purchased from Sinopharm Chemical Reagent Co., Ltd.. RTIL [EMIM][$BF_4$] (Lanzhou Yulu Fine Chemical Co., Ltd) was purified via the Schlenk line at 85°C for 48 h.

Based on synthesis strategies reported in the previous work[60], we synthesized $Ni_3(HITP)_2$ with modifications for better crystallinity. 120 mL of the stock solution (2.33 mmol·$L^{-1}$) of nickel(II) acetate tetrahydrate ($Ni(OAc)_2·4H_2O$) in DMF/DMA (v/v = 1:1) was preheated to 65°C via the oil bath in a 500 mL round bottom flask, to which was added 80 mL of fresh NaOAc aqueous solution (2 mol·$L^{-1}$) under agitation. The above solution was heated to 65°C before a solution of 100 mg of HATP·6HCl in 30 mL of water was added, and this mixture was heated with high-speed stirring for 4 hours at 65°C. Afterward, the suspension was cooled to room temperature. The supernatant solution was removed before the resulting black precipitate was centrifuged, and the solvent was exchanged with water and methanol. Finally, the residue was put under vacuum.

## Acknowledgements

The authors acknowledge the funding support from the National Natural Science Foundation of China (T2325012, 92472109), the Program for HUST Academic Frontier Youth Team, and Wuhan Supercomputing Center for providing computational resources. Liang Zeng is supported by the Postdoctoral Fellowship Program of CPSF (GZC20240532).

# References

1.  Simon, P.; Gogotsi, Y.; Dunn, B., Where do batteries end and supercapacitors begin? *Science* **2014,** *343* (6176), 1210-1211.

2.  Keum, K.; Kim, J. W.; Hong, S. Y.; Son, J. G.; Lee, S. S.; Ha, J. S., Flexible/stretchable supercapacitors with novel functionality for wearable electronics. *Adv. Mater.* **2020,** *32* (51), 2002180.

3.  Pomerantseva, E.; Bonaccorso, F.; Feng, X.; Cui, Y.; Gogotsi, Y., Energy storage: The future enabled by nanomaterials. *Science* **2019,** *366* (6468), eaan8285.

4.  Simon, P.; Gogotsi, Y., Perspectives for electrochemical capacitors and related devices. *Nat. Mater.* **2020,** *19* (11), 1151-1163.

5.  Xie, K.; Qin, X.; Wang, X.; Wang, Y.; Tao, H.; Wu, Q.; Yang, L.; Hu, Z., Carbon nanocages as supercapacitor electrode materials. *Adv. Mater.* **2011,** *24* (3), 347-352.

6.  Li, H.; Tao, Y.; Zheng, X.; Luo, J.; Kang, F.; Cheng, H.-M.; Yang, Q.-H., Ultra-thick graphene bulk supercapacitor electrodes for compact energy storage. *Energy Environ. Sci.* **2016,** *9* (10), 3135-3142.

7.  Mo, T.; Wang, Z.; Zeng, L.; Chen, M.; Kornyshev, A. A.; Zhang, M.; Zhao, Y.; Feng, G., Energy storage mechanism in supercapacitors with porous graphdiynes: Effects of pore topology and electrode metallicity. *Adv. Mater.* **2023,** *35* (33), 2301118.

8.  Barbieri, O.; Hahn, M.; Herzog, A.; Kötz, R., Capacitance limits of high surface area activated carbons for double layer capacitors. *Carbon* **2005,** *43* (6), 1303-1310.

9.  Chmiola, J.; Yushin, G.; Gogotsi, Y.; Portet, C.; Simon, P.; Taberna, P.-L., Anomalous increase in carbon capacitance at pore sizes less than 1 nanometer. *Science* **2006,** *313* (5794), 1760-1763.

10. Fei, H.-F.; Li, W.; Bhardwaj, A.; Nuguri, S.; Ribbe, A.; Watkins, J. J., Ordered nanoporous carbons with broadly tunable pore size using bottlebrush block copolymer templates. *J. Am. Chem. Soc.* **2019,** *141* (42), 17006-17014.

11. Wu, J., Understanding the Electric Double-Layer Structure, Capacitance, and Charging Dynamics. *Chem. Rev.* **2022,** *122* (12), 10821-10859.

12. Liu, Y. M.; Merlet, C.; Smit, B., Carbons with regular pore geometry yield fundamental insights into supercapacitor charge storage. *ACS Cent. Sci.* **2019,** *5* (11), 1813-1823.

13. Liu, X.; Lyu, D.; Merlet, C.; Leesmith, M. J.; Hua, X.; Xu, Z.; Grey, C. P.; Forse, A. C., Structural disorder determines capacitance in nanoporous carbons. *Science* **2024,** *384* (6693), 321-325.

14. Forse, A. C.; Merlet, C.; Griffin, J. M.; Grey, C. P., New perspectives on the charging mechanisms of supercapacitors. *J. Am. Chem. Soc.* **2016,** *138* (18), 5731-5744.

15. Jäckel, N.; Rodner, M.; Schreiber, A.; Jeongwook, J.; Zeiger, M.; Aslan, M.; Weingarth, D.; Presser, V., Anomalous or regular capacitance? The influence of pore size dispersity on double-layer formation. *J. Power Sources* **2016,** *326*, 660-671.

16. Jablonka, K. M.; Ongari, D.; Moosavi, S. M.; Smit, B., Big-data science in porous materials: materials genomics and machine learning. *Chem. Rev.* **2020,** *120* (16), 8066-8129.

17. Wang, Z.; Sun, Z.; Yin, H.; Liu, X.; Wang, J.; Zhao, H.; Pang, C. H.; Wu, T.; Li, S.; Yin, Z., Data-Driven Materials Innovation and Applications. *Adv. Mater.* **2022,** *34* (36), 2104113.

18. Su, H.; Lin, S.; Deng, S.; Lian, C.; Shang, Y.; Liu, H., Predicting the capacitance of carbon-based electric double layer capacitors by machine learning. *Nanoscale Adv.* **2019,** *1* (6), 2162-2166.

19. Zhou, M.; Gallegos, A.; Liu, K.; Dai, S.; Wu, J., Insights from machine learning of carbon electrodes for electric double layer capacitors. *Carbon* **2020,** *157*, 147-152.

20. Liu, P.; Wen, Y.; Huang, L.; Zhu, X.; Wu, R.; Ai, S.; Xue, T.; Ge, Y., An emerging machine learning strategy for the assisted-design of high-performance supercapacitor materials by mining the relationship between capacitance and structural features of porous carbon. *J. Electroanal. Chem.* **2021,** *899*, 115684.

21. Tawfik, W. Z.; Mohammad, S. N.; Rahouma, K. H.; Tammam, E.; Salama, G. M., An artificial neural network model for capacitance prediction of porous carbon-based supercapacitor electrodes. *J. Energy Storage* **2023,** *73*, 108830.

22. Wang, T.; Pan, R.; Martins, M. L.; Cui, J.; Huang, Z.; Thapaliya, B. P.; Do-Thanh, C.-L.; Zhou, M.; Fan, J.; Yang, Z., Machine-learning-assisted material discovery of oxygen-rich highly porous carbon active materials for aqueous supercapacitors. *Nat. Commun.* **2023,** *14* (1), 4607.

23. Salanne, M.; Rotenberg, B.; Naoi, K.; Kaneko, K.; Taberna, P.-L.; Grey, C. P.; Dunn, B.; Simon, P., Efficient storage mechanisms for building better supercapacitors. *Nat. Energy* **2016,** *1* (6), 1-10.

24. Jackel, N.; Simon, P.; Gogotsi, Y.; Presser, V., Increase in capacitance by subnanometer pores in carbon. *ACS Energy Lett.* **2016,** *1* (6), 1262-1265.

25. Walton, K. S.; Snurr, R. Q., Applicability of the BET method for determining surface areas of microporous metal− organic frameworks. *J. Am. Chem. Soc.* **2007,** *129* (27), 8552-8556.

26. Balhatchet, C. J.; Gittins, J. W.; Shin, S.-J.; Ge, K.; Liu, X.; Trisukhon, T.; Sharma, S.; Kress, T.; Taberna, P.-L.; Simon, P., Revealing Ion Adsorption and Charging Mechanisms in Layered Metal–Organic Framework Supercapacitors with Solid-State Nuclear Magnetic Resonance. *J. Am. Chem. Soc.* **2024,** *146* (33), 23171-23181.

27. Rosen, A. S.; Iyer, S. M.; Ray, D.; Yao, Z.; Aspuru-Guzik, A.; Gagliardi, L.; Notestein, J. M.; Snurr, R. Q., Machine learning the quantum-chemical properties of metal–organic frameworks for accelerated materials discovery. *Matter* **2021,** *4* (5), 1578-1597.

28. Lin, J.; Zhang, H.; Asadi, M.; Zhao, K.; Yang, L.; Fan, Y.; Zhu, J.; Liu, Q.; Sun, L.; Xie, W. J., Machine Learning-Driven Discovery and Structure–Activity Relationship Analysis of Conductive Metal–Organic Frameworks. *Chem. Mater.* **2024**.

29. Merlet, C.; Rotenberg, B.; Madden, P. A.; Taberna, P.-L.; Simon, P.; Gogotsi, Y.; Salanne, M., On the molecular origin of supercapacitance in nanoporous carbon electrodes. *Nat. Mater.* **2012,** *11* (4), 306-310.

30. Breitsprecher, K.; Janssen, M.; Srimuk, P.; Mehdi, B. L.; Presser, V.; Holm, C.; Kondrat, S., How to speed up ion transport in nanopores. *Nat. Commun.* **2020,** *11* (1), 6085.

31. Bi, S.; Banda, H.; Chen, M.; Niu, L.; Chen, M.; Wu, T.; Wang, J.; Wang, R.; Feng, J.; Chen, T.; Dinca, M.; Kornyshev, A. A.; Feng, G., Molecular understanding of charge storage and charging dynamics in supercapacitors with MOF electrodes and ionic liquid electrolytes. *Nat. Mater.* **2020,** *19* (5), 552-558.

32. Rosen, A. S.; Fung, V.; Huck, P.; O'Donnell, C. T.; Horton, M. K.; Truhlar, D. G.; Persson, K. A.; Notestein, J. M.; Snurr, R. Q., High-throughput predictions of metal–organic framework electronic properties: theoretical challenges, graph neural networks, and data exploration. *npj Comput. Mater.* **2022,** *8* (1), 1-10.

33. Willems, T. F.; Rycroft, C. H.; Kazi, M.; Meza, J. C.; Haranczyk, M., Algorithms and tools for high-throughput geometry-based analysis of crystalline porous materials. *Microporous Mesoporous Mater.* **2012,** *149* (1), 134-141.

34. Feng, D.; Lei, T.; Lukatskaya, M. R.; Park, J.; Huang, Z.; Lee, M.; Shaw, L.; Chen, S.; Yakovenko, A. A.; Kulkarni, A., Robust and conductive two-dimensional metal− organic frameworks with exceptionally high volumetric and areal capacitance. *Nat. Energy* **2018,** *3* (1), 30-36.

35. Talin, A. A.; Centrone, A.; Ford, A. C.; Foster, M. E.; Stavila, V.; Haney, P.; Kinney, R. A.; Szalai, V.; El Gabaly, F.; Yoon, H. P., Tunable electrical conductivity in metal-organic framework thin-film devices. *Science* **2014,** *343* (6166), 66-69.

36. Park, G.; Demuth, M. C.; Hendon, C. H.; Park, S. S., Acid-Dependent Charge Transport in a Solution-Processed 2D Conductive Metal-Organic Framework. *J. Am. Chem. Soc.* **2024,** *146* (16), 11493-11499.

37. Asuero, A. G.; Sayago, A.; González, A., The correlation coefficient: An overview. *Crit. Rev. Anal. Chem.* **2006,** *36* (1), 41-59.

38. Greenacre, M.; Groenen, P. J.; Hastie, T.; d'Enza, A. I.; Markos, A.; Tuzhilina, E., Principal component analysis. *Nat. Rev. Methods Primers* **2022,** *2* (1), 100.

39. Chen, T.; Guestrin, C., Xgboost: A scalable tree boosting system. *Proceedings of the 22nd ACM SIGKDD International Conference on Knowledge Discovery and Data Mining* **2016**, 785-794.

40. Shapley, L. S., A value for n-person games. *RAND Corporation* **1953**.

41. Chen, H.; Covert, I. C.; Lundberg, S. M.; Lee, S.-I., Algorithms to estimate Shapley value feature attributions. *Nat. Mach. Intell.* **2023,** *5* (6), 590-601.

42. Sheberla, D.; Bachman, J. C.; Elias, J. S.; Sun, C. J.; Shao-Horn, Y.; Dinca, M., Conductive MOF electrodes for stable supercapacitors with high areal capacitance. *Nat. Mater.* **2017,** *16* (2), 220-224.

43. Wang, Q.; Yan, J.; Fan, Z., Carbon materials for high volumetric performance supercapacitors: design, progress, challenges and opportunities. *Energy Environ. Sci.* **2016,** *9* (3), 729-762.

44. Mo, T.; Bi, S.; Zhang, Y.; Presser, V.; Wang, X.; Gogotsi, Y.; Feng, G., Ion structure transition enhances charging dynamics in subnanometer pores. *ACS Nano* **2020,** *14* (2), 2395-2403.

45. Tivony, R.; Safran, S.; Pincus, P.; Silbert, G.; Klein, J., Charging dynamics of an individual nanopore. *Nat. Commun.* **2018,** *9* (1), 4203.

46. Kondrat, S.; Wu, P.; Qiao, R.; Kornyshev, A. A., Accelerating charging dynamics in subnanometre pores. *Nat. Mater.* **2014,** *13* (4), 387-93.

47. Mo, T.; Zhou, J.; He, H.; Zhu, B., Oscillation Charging Dynamics in Nanopore Supercapacitors with Organic Electrolyte. *ACS Appl. Mater. Interfaces* **2023,** *15* (44), 51274-51280.

48. Li, Z.; Misra, R. P.; Li, Y.; Yao, Y.-C.; Zhao, S.; Zhang, Y.; Chen, Y.; Blankschtein, D.; Noy, A., Breakdown of the Nernst–Einstein relation in carbon nanotube porins. *Nat. Nanotechnol.* **2023,** *18* (2), 177-183.

49. Xu, J.; He, Y.; Bi, S.; Wang, M.; Yang, P.; Wu, D.; Wang, J.; Zhang, F., An olefin-linked covalent organic framework as a flexible thin-film electrode for a high-performance micro-supercapacitor. *Angew. Chem,* **2019,** *131* (35), 12193-12197.

50. Mostazo-López, M. J.; Ruiz-Rosas, R.; Castro-Muñiz, A.; Nishihara, H.; Kyotani, T.; Morallón, E.; Cazorla-Amorós, D., Ultraporous nitrogen-doped zeolite-templated carbon for high power density aqueous-based supercapacitors. *Carbon* **2018,** *129*, 510-519.

51. De Villenoisy, T.; Zheng, X.; Wong, V.; Mofarah, S. S.; Arandiyan, H.; Yamauchi, Y.; Koshy, P.; Sorrell, C. C., Principles of design and synthesis of metal derivatives from MOFs. *Adv. Mater.* **2023,** *35* (24), 2210166.

52. Kang, Y.; Park, H.; Smit, B.; Kim, J., A multi-modal pre-training transformer for universal transfer learning in metal–organic frameworks. *Nat. Mach. Intell.* **2023,** *5* (3), 309-318.

53. Rappé, A. K.; Casewit, C. J.; Colwell, K.; Goddard III, W. A.; Skiff, W. M., UFF, a full periodic table force field for molecular mechanics and molecular dynamics simulations. *J. Am. Chem. Soc.* **1992,** *114* (25), 10024-10035.

54. Doherty, B.; Zhong, X.; Acevedo, O., Virtual site OPLS force field for imidazolium-based ionic liquids. *J. Phys. Chem. B.* **2018,** *122* (11), 2962-2974.

55. Hess, B.; Kutzner, C.; Van Der Spoel, D.; Lindahl, E., GROMACS 4: algorithms for highly efficient, load-balanced, and scalable molecular simulation. *J. Chem. Theory Comput.* **2008,** *4* (3), 435-447.

56. Siepmann, J. I.; Sprik, M., Influence of surface topology and electrostatic potential on water/electrode systems. *J. Chem. Phys.* **1998,** *102* (1), 511.

57. Zeng, L.; Wu, T.; Ye, T.; Mo, T.; Qiao, R.; Feng, G., Modeling galvanostatic charge–discharge of nanoporous supercapacitors. *Nat. Comput. Sci.* **2021,** *1* (11), 725-731.

58. Bussi, G.; Donadio, D.; Parrinello, M., Canonical sampling through velocity rescaling. *J. Chem. Phys.* **2007,** *126* (1), 014101.

59. Gingrich, T. R.; Wilson, M., On the Ewald summation of Gaussian charges for the simulation of metallic surfaces. *Chem. Phys. Lett.* **2010,** *500* (1-3), 178-183.

60. Chen, T.; Dou, J.-H.; Yang, L.; Sun, C.; Libretto, N. J.; Skorupskii, G.; Miller, J. T.; Dincă, M., Continuous electrical conductivity variation in M3 (hexaiminotriphenylene) 2 (M= Co, Ni, Cu) MOF alloys. *J. Am. Chem. Soc.* **2020,** *142* (28), 12367-12373.